# FROM LOCAL DEFECTS TO SHEAR-ORGANIZED BIOFILMS IN TONSILLAR CRYPTS VIA COMPUTATIONAL SIMULATIONS

Arturo Tozzi (corresponding author)
ASL Napoli 1 Centro, Distretto 27, Naples, Italy
Via Comunale del Principe 13/a 80145
tozziarturo@libero.it

## ABSTRACT

Biofilms in human tonsillar crypts show long-term persistence with episodic dispersal that current biochemical and microbiological descriptions do not fully explain, particularly with respect to spatial localization. We introduce a biophysical framework in which tonsillar biofilm dynamics arise from the interaction between two mechanical phenomena: a Kosterlitz–Thouless–type defect nucleation process and a Kelvin–Helmholtz–type shear-driven interfacial instability. Crypt geometry is modeled as a confined, heterogeneous environment that promotes mechanically persistent surface defects generated by growth-induced compression. Tangential shear associated with breathing and swallowing selectively amplifies these defects, producing organized surface deformations. Numerical simulations show that only the coexistence of both mechanisms yields localized, propagating, and persistent interface structures, whereas their absence leads to diffuse, unstructured dynamics.

INTRODUCTION

Biofilms colonizing the human tonsillar crypts form a mechanically complex system in which long-term persistence coexists with episodic reorganization and dispersal. Investigations have primarily addressed microbiological composition, extracellular polymeric substances, immune interactions and chemical gradients, providing detailed insight into biochemical regulation and ecological stability (Zijnge et al. 2010; Washio and Takahashi 2016; Chevalier, Ranque, and Prêcheur 2018; Brown et al. 2019; Bugari et al. 2021; Thurnheer and Paqué 2021; Das et al. 2023; Brown et al. 2023; Bloch et al. 2024). Mechanical aspects are usually incorporated in simplified terms, such as bulk viscosity, adhesion strength or erosion by flow and are often treated as secondary modifiers of biologically driven processes (Stojković et al. 2015; Hart et al. 2019; Boyd et al. 2021; Wang et al. 2022; Li et al. 2023; Waldman et al. 2023; Teixeira et al. 2023). Although computational and experimental studies have demonstrated that biofilms exposed to flow can develop ripple-like surface deformations consistent with shear-driven interfacial instabilities (Cogan et al. 2018), these approaches assume spatially homogeneous material properties and do not explicitly address the source of localized surface irregularities. As a consequence, existing models struggle to account for the stable, spatially anchored structures observed within tonsillar crypts under low or intermittent shear, as well as for the abrupt transition to organized surface deformation during coughing, swallowing or speech. These purely hydrodynamic or uniformly elastic descriptions cannot capture the combined effects of crypt geometry, growth-induced stress and material heterogeneity, motivating the need for a framework to distinguish between mechanisms responsible for defect nucleation and those responsible for shear-driven amplification.

We introduce a mechanistic framework in which tonsillar biofilm dynamics emerge from the interaction of two distinct physical mechanisms: a Kosterlitz–Thouless–type defect nucleation process and a Kelvin–Helmholtz–type shear-driven interfacial instability. The first mechanism treats localized surface irregularities as mechanically persistent defects arising from growth-induced compression and spatial heterogeneity, analogous to unbound defects in a Kosterlitz–Thouless transition in two-dimensional systems (Martin et al. 2017; Kim et al. 2019; Chen, Srolovitz, and Han 2020; Veyrat et al. 2023; Aguilera et al. 2024). The second mechanism captures the amplification of pre-existing surface perturbations under tangential shear, in line with Kelvin–Helmholtz–type instabilities identified in fluid interfaces under flow (Soloviev et al. 2017; Caspary et al. 2018; Han et al. 2021; Ge et al. 2022; Stawarz et al. 2024). By explicitly coupling these two mechanisms, our framework separates the conditions governing defect formation from those governing their dynamic organization. We implement this coupling in a minimal computational model that allows comparison between a regime in which both mechanisms are active and a regime lacking both. The resulting simulations generate structured space–time patterns, growth-cone dynamics and scalar localization measures to characterize how crypt geometry could partition biofilm behavior into nucleation-dominated and shear-organized regions.

We will proceed as follows. The Methods section introduces the geometric abstraction, mathematical formulation and numerical implementation of our model. Subsequent sections present simulation results and quantitative comparisons between dynamical regimes, followed by a discussion situating the findings within the existing literature on biofilm mechanics and interfacial instabilities.

METHODS

We present here the geometric abstraction, mathematical formulation, numerical implementation and analytical procedures to assess tonsillar biofilm dynamics governed by the interaction between Kosterlitz–Thouless–type defect nucleation and Kelvin–Helmholtz–type shear-driven interfacial instability.

**Geometric abstraction of the tonsillar crypt**. Our geometric abstraction aims to define the spatial domain in which the two mechanisms operate and interact. The tonsillar crypt is represented as a confined surface region corresponding to the crypt mouth, where tangential shear from airflow and saliva is maximal and where interface deformation is most pronounced. The geometry is reduced to a one-dimensional coordinate $x \in [0, L]$, aligned with the circumferential direction of the crypt opening. This reduction captures lateral organization along the rim while treating crypt depth implicitly through boundary conditions and spatial heterogeneity. The biofilm is modeled as a thin deformable layer adhering to a compliant substrate representing the crypt epithelium. Spatial heterogeneity arising from growth history, extracellular polymeric substance distribution and epithelial microstructure is encoded directly into the surface properties rather than modeled as a separate field. The spatial domain is discretized uniformly with spacing $\Delta x = L/N$ and no-flux boundary conditions are applied at the domain edges. This geometric abstraction, schematically illustrated in Fig. 1, allows separation between regions dominated by defect nucleation and regions dominated by shear-driven amplification.

**Kosterlitz–Thouless–type defect nucleation field**. The localized surface heterogeneity is here formalized as a defect field inspired by Kosterlitz–Thouless physics. The first mechanism is a Kosterlitz–Thouless–type defect nucleation process, introduced to represent the emergence of stable, localized surface irregularities generated by growth-induced compression and material disorder. Rather than modeling defect unbinding dynamically, the post-nucleation defect landscape is treated as a static field $\rho(x)$, representing the spatial density of unbound defects. Discrete defect cores are placed at positions $x_i$, $i = 1, \ldots, N_d$ and each core contributes a localized influence described by a Gaussian kernel. The resulting defect field is defined as

$$\rho(x) = \frac{1}{\rho_{\max}} \sum_{i=1}^{N_d} \exp\left(-\frac{(x-x_i)^2}{2\sigma^2}\right),$$

where $\sigma$ controls the spatial extent of defect influence and $\rho_{\max}$ normalizes the field. This construction reflects the physical assumption that, following a Kosterlitz–Thouless–type unbinding transition, defects are discrete, persistent and spatially localized. The field $\rho(x)$ encodes the mechanical memory of the biofilm surface and remains fixed throughout the simulation, allowing isolation of the second mechanism.

**Kelvin–Helmholtz–type shear representation**. Shear forcing is now introduced as the driver of Kelvin–Helmholtz–type interfacial instability. The second mechanism is a Kelvin–Helmholtz–type instability induced by tangential shear between the biofilm surface and the surrounding fluid. Shear is represented by a dimensionless parameter $U$, proportional to the velocity difference between the fluid phase and the biofilm surface. Rather than explicitly solving Navier–Stokes equations, shear enters the model as a linear amplification term acting on interface perturbations, consistent with reduced descriptions of Kelvin–Helmholtz instability near onset. The parameter $U$ is held constant during each simulation and varied across runs to probe different dynamical regimes. This formulation captures the essential role of shear as a destabilizing influence without introducing unnecessary hydrodynamic complexity.

**Interface height field and governing equation**. We show here that the coupled action of the two mechanisms can be expressed through a nonlinear evolution equation for the interface. The biofilm–fluid interface is described by a scalar height field $h(x, t)$, measuring normal displacement from a reference surface. Its evolution is governed by

$$\frac{\partial h}{\partial t} = (\alpha U + \beta \rho(x))h - \gamma h^3 - \kappa \frac{\partial^2 h}{\partial x^2}.$$

The term $\alpha U h$ represents uniform shear-driven amplification associated with Kelvin–Helmholtz–type instability, while $\beta \rho(x) h$ represents defect-localized amplification associated with Kosterlitz–Thouless–type defect nucleation. The cubic saturation term $-\gamma h^3$, with $\gamma > 0$, ensures bounded growth and represents nonlinear elastic or geometric stiffening. The smoothing term $-\kappa \partial_x^2 h$ represents effective interfacial tension, bending resistance or elastic regularization. This equation constitutes a minimal normal form capturing the interaction between localized defects and shear-driven instability.

**Numerical discretization and integration scheme**. The continuous model can now be converted into a discrete dynamical system using explicit numerical schemes. Spatial derivatives are approximated using second-order central finite differences. Time integration is performed using an explicit Euler method with time step $\Delta t$. The discrete update equation is

$$h_j^{n+1} = h_j^n + \Delta t \left[ (\alpha U + \beta \rho_j) h_j^n - \gamma (h_j^n)^3 - \kappa \frac{h_{j+1}^n - 2h_j^n + h_{j-1}^n}{(\Delta x)^2} \right],$$

where j indexes spatial position and n indexes time. Initial conditions are small-amplitude random perturbations $h(x, 0) \sim \mathcal{N}(0, \epsilon^2)$, ensuring unbiased symmetry breaking. Parameters are chosen to satisfy numerical stability constraints and to allow observation of both transient growth and saturated dynamics.

**Space–time field construction**. The global spatiotemporal organization is analyzed here through space–time representations. The full interface field $h(x, t)$ is recorded at each time step, producing a matrix $H_{n,j}$. Space–time maps are constructed by plotting H as a color-coded image with axes x and t. To facilitate comparison across runs, the field is normalized by the standard deviation of early-time fluctuations. These maps reveal whether amplification remains localized, spreads laterally or remains diffuse, providing direct visual discrimination between dynamical regimes.

**Growth cone extraction procedure**. Then, the propagation of deformation is quantified through growth cone analysis. To extract growth cones, a threshold $\theta$ is defined based on baseline noise, $\theta = k\sigma_0$, where $\sigma_0$ is the median absolute deviation of h during an initial time window. For each spatial location $x_j$, the arrival time $t_j$ is defined as

$$t_j = \min \{t: | h(x_j, t) | > \theta\}.$$

The locus $(x_j, t_j)$ defines a growth front. Growth cones are visualized as contour lines superimposed on space–time maps. To reduce noise, lower-envelope extraction is applied by binning distances from defect centers and selecting minimal arrival times per bin.

**Localization index and scalar metrics**. Spatial concentration of deformation is then quantified using a scalar order parameter.
A localization index $\Lambda(t)$ is defined as

$$\Lambda(t) = \frac{\text{std}_x[h(x, t)]}{\langle | h(x, t) | \rangle_x},$$

where $\text{std}_x$ is the spatial standard deviation and $\langle \cdot \rangle_x$ denotes spatial averaging. High values of $\Lambda$ indicate deformation concentrated at specific locations, while low values indicate diffuse fluctuations. This index provides a compact quantitative discriminator between organized and unorganized regimes.

**Comparison of two dynamical regimes**. Two regimes are simulated. In the first, both Kosterlitz–Thouless–type defect nucleation and Kelvin–Helmholtz–type shear instability are active, corresponding to $\beta > 0$ and $U > 0$. In the second, both mechanisms are suppressed by setting $\beta = 0$ and reducing U below instability threshold. All other parameters, initial conditions and numerical procedures are identical. Differences in space–time structure, growth cones and localization indices could therefore be attributable solely to the presence or absence of the two mechanisms.

**Computational tools and reproducibility**. All computations are implemented in Python, using NumPy for numerical operations and Matplotlib for visualization. Random seeds are fixed for reproducibility. No external data are used and all parameters are explicitly specified within the simulation code.

In conclusion, we specified here the mathematical and computational framework for studying tonsillar biofilm dynamics governed by Kosterlitz–Thouless–type defect nucleation and Kelvin–Helmholtz–type shear-driven instability, enabling reproducible analysis and controlled comparison between distinct mechanical regimes.

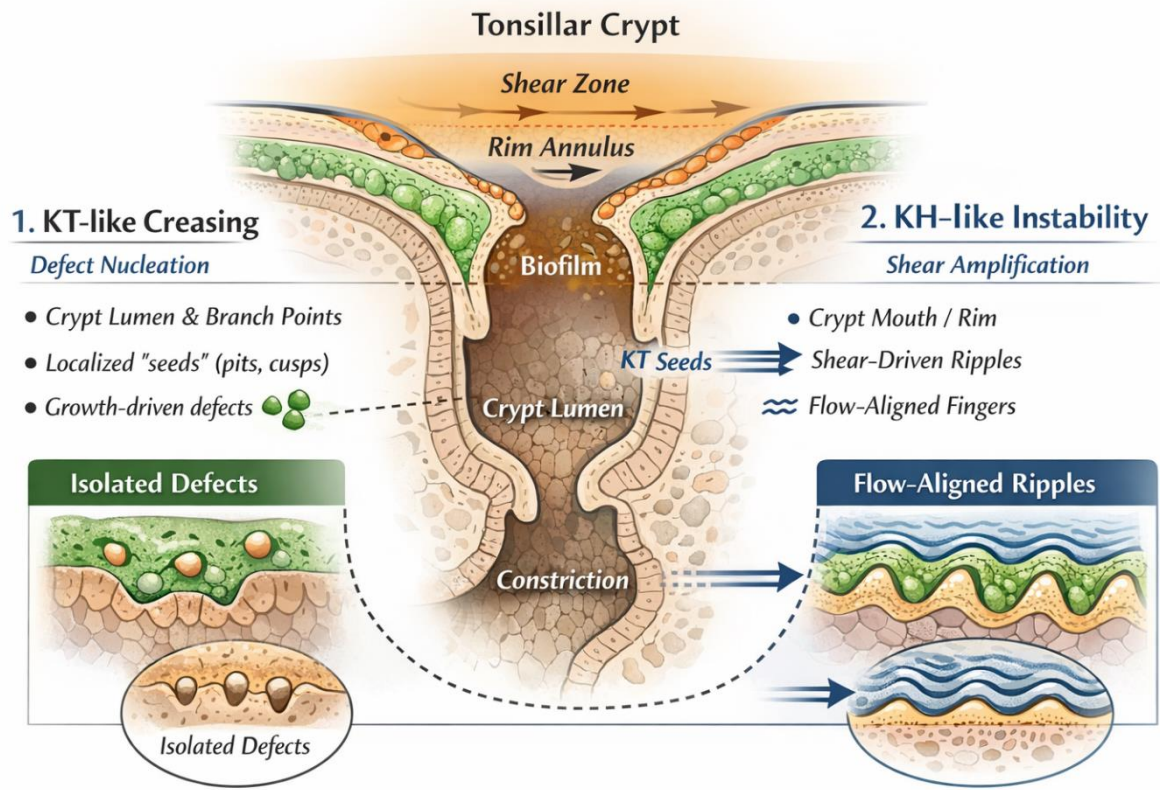

**Figure 1**. Conceptual framework and geometric setting of a two-stage mechanical organization of biofilms in tonsillar crypts. Growth, confinement and material heterogeneity within the crypt lumen and branch points generate localized surface defects (pits or cusps), interpreted as a KT-like creasing process producing stable nucleation seeds largely insensitive to flow. At the crypt mouth and rim annulus, strong tangential shear from airflow or saliva acts on the biofilm–fluid interface. Pre-existing defects are selectively amplified into flow-aligned ripples and finger-like protrusions via a KH-like instability.

RESULTS

The outcomes of the numerical simulations are reported here, focusing on the spatiotemporal organization of the biofilm–fluid interface, the emergence and propagation of localized deformations and the quantitative comparison between dynamical regimes with and without the two mechanisms.

**Defect-seeded amplification and spatial organization**. Simulations including both Kosterlitz–Thouless–type defect nucleation and Kelvin–Helmholtz–type shear-driven instability produced spatially heterogeneous interface dynamics characterized by persistent localization. In the coupled regime, interface deformation remained bounded yet nonuniform, with peak amplitudes systematically emerging in the vicinity of defect sites (Figure 2). The final interface profiles showed correspondence between the spatial distribution of the defect field and the positions of maximal deformation, while defect-free regions exhibited markedly lower amplitudes. Space–time representations revealed that amplification initiated at discrete locations and subsequently spread laterally, forming elongated deformation bands aligned with the crypt mouth coordinate (Figure 3). Growth-cone analysis confirmed this behavior quantitatively: for each defect, the arrival time to threshold increased approximately monotonically with distance, yielding well-defined cone fronts whose slopes varied modestly across seeds (Figure 4). Across simulations, the lateral extent of these cones was limited by nonlinear saturation and interfacial smoothing, preventing global destabilization. In contrast, early-time fluctuations away from defect sites decayed or remained near baseline levels.
These results establish that, under constant shear, localized defects act as organizing centers that structure interface evolution in both space and time, providing a dynamical signature of the coupled mechanism and setting the basis for regime-level comparison.

**Comparison between dynamical regimes**. Direct comparison between the coupled regime and a regime lacking both defect nucleation and shear-driven instability revealed pronounced differences. In the absence of the two mechanisms, interface fluctuations remained spatially diffuse, with no persistent anchoring to specific locations and no coherent lateral propagation visible in space–time maps (Figure 3). To quantify these differences, a localization index was computed over time for both regimes. In the coupled regime, the index increased rapidly from baseline values to a distinct peak, reflecting the concentration of deformation around defect sites and then relaxed toward a steady level as patterns saturated. In the uncoupled regime, the localization index remained approximately constant over time, indicating the absence of spatial concentration (Figure 5). This contrast was consistent across repeated simulations with different random initial perturbations. The tabulated comparison summarizes these observations, highlighting differences in onset behavior, spatial memory, propagation features and temporal variability between regimes (Table).

Together, these measurements suggest that the inclusion of both Kosterlitz–Thouless–type defect nucleation and Kelvin–Helmholtz–type shear amplification produces a distinct class of dynamics that cannot be reproduced by uniform growth and smoothing alone.

Overall, our results show that defect-seeded shear amplification yields localized, propagating and persistent interface deformations, whereas simulations lacking both mechanisms display diffuse and unstructured dynamics. Growth-cone extraction and localization metrics consistently distinguish the two regimes. These findings quantitatively characterize how localized defects and shear jointly organize biofilm surface dynamics within the modeled crypt geometry.

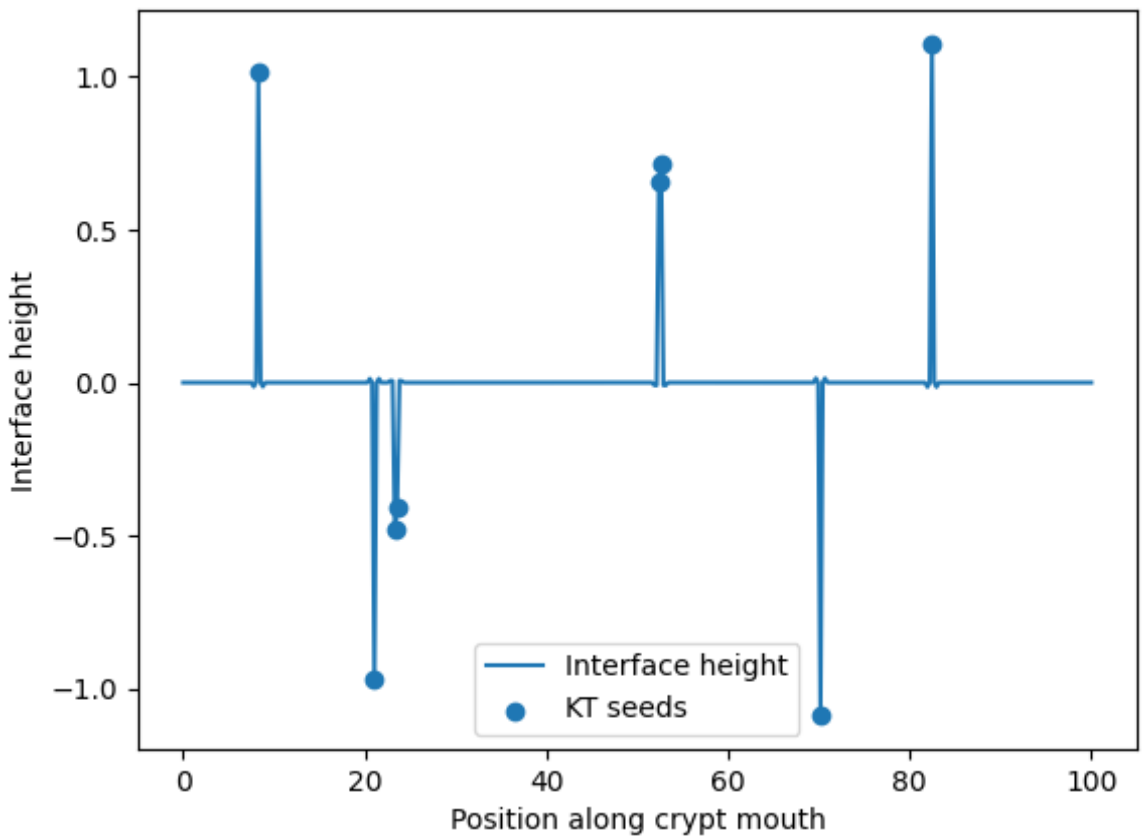

**Figure 2**. Coupled simulation of defect-seeded shear amplification along the crypt mouth. Discrete defect sites, generated by the KT-like nucleation stage, locally enhance instability growth under shear. Interface deformation preferentially amplifies near defect locations, producing localized peaks and troughs acting as precursors to ripple formation and finger-like protrusions. <u>Therefore, crypt geometry and defect localization bias shear-driven pattern selection and potential fragment release sites.</u>

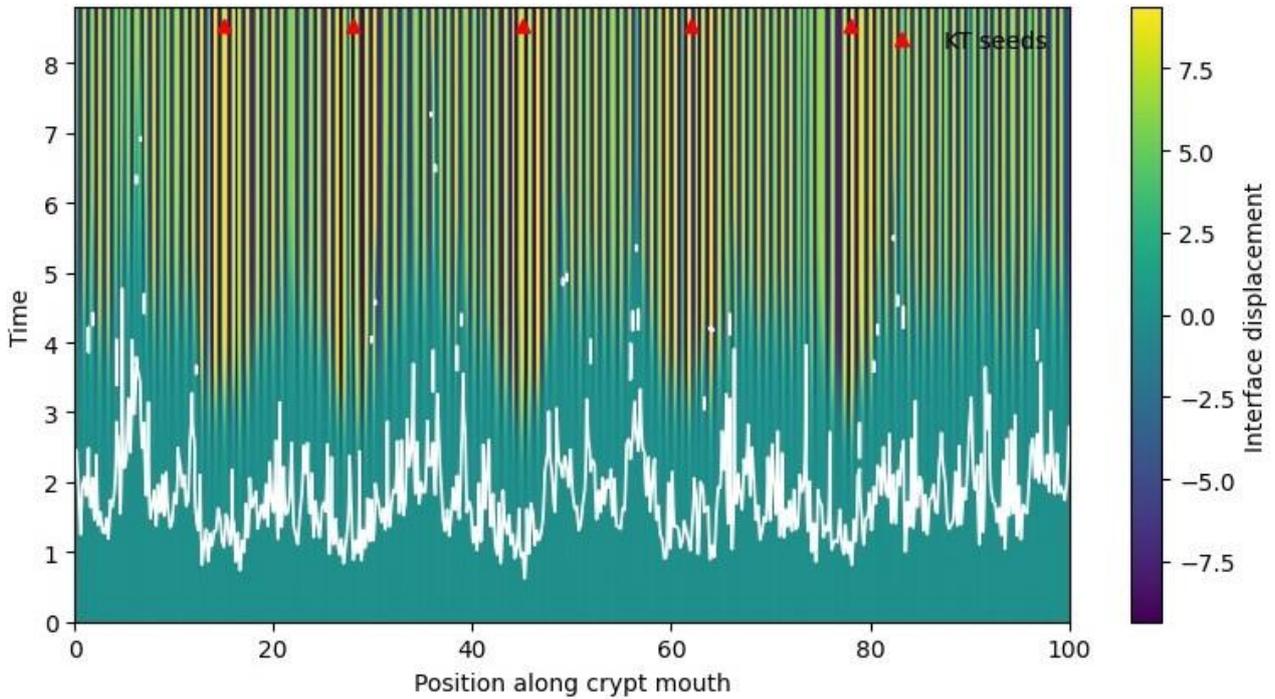

**Figure 3**. Space–time evolution of the biofilm–fluid interface along the crypt mouth under constant shear in the presence of localized KT-like defect seeds (red triangles). Color encodes interface displacement as a function of position and time. Amplification initiates preferentially at defect locations and subsequently spreads laterally, while defect-free regions remain weakly perturbed. White contours indicate growth-cone boundaries, defined as the earliest threshold-crossing of interface displacement above the baseline noise level. The figure suggests that pre-existing surface heterogeneities bias shear-driven instability, transforming localized nucleation sites into extended, time-persistent deformation bands at the crypt mouth.

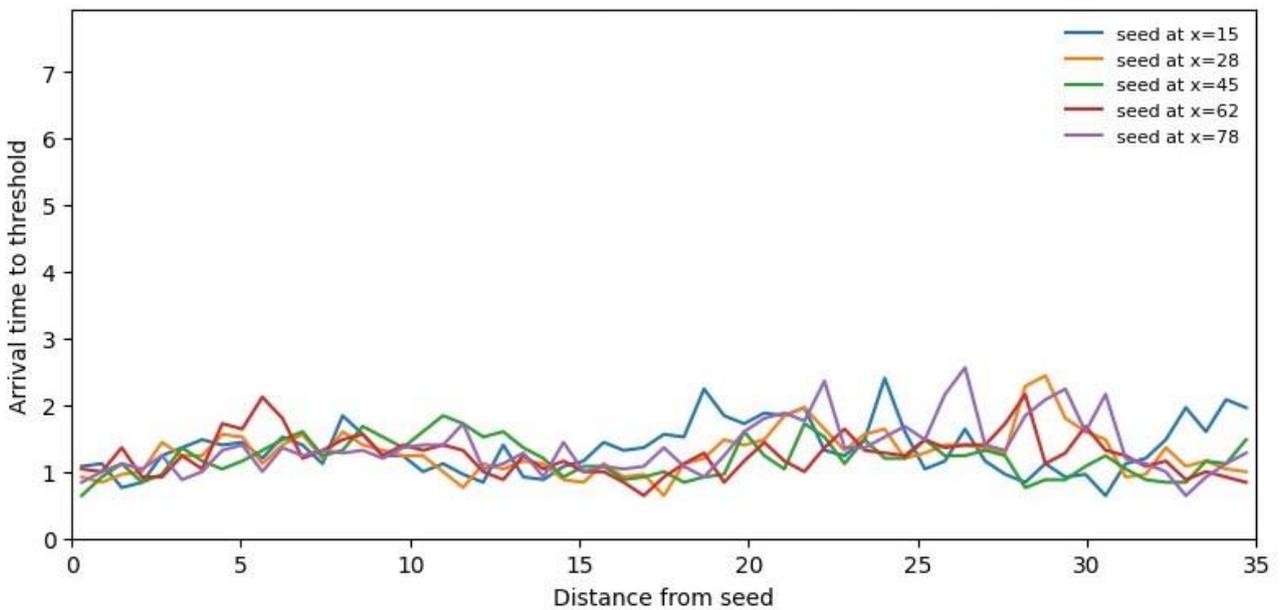

**Figure 4**. Growth-cone fronts quantified as arrival time to threshold versus distance from each defect seed. For each seed, the earliest threshold-crossing time is computed across space and summarized by a lower-envelope front, yielding a cone-like time–distance relation. The slope provides an operational estimate of lateral propagation speed of defect-seeded shear amplification along the crypt mouth.

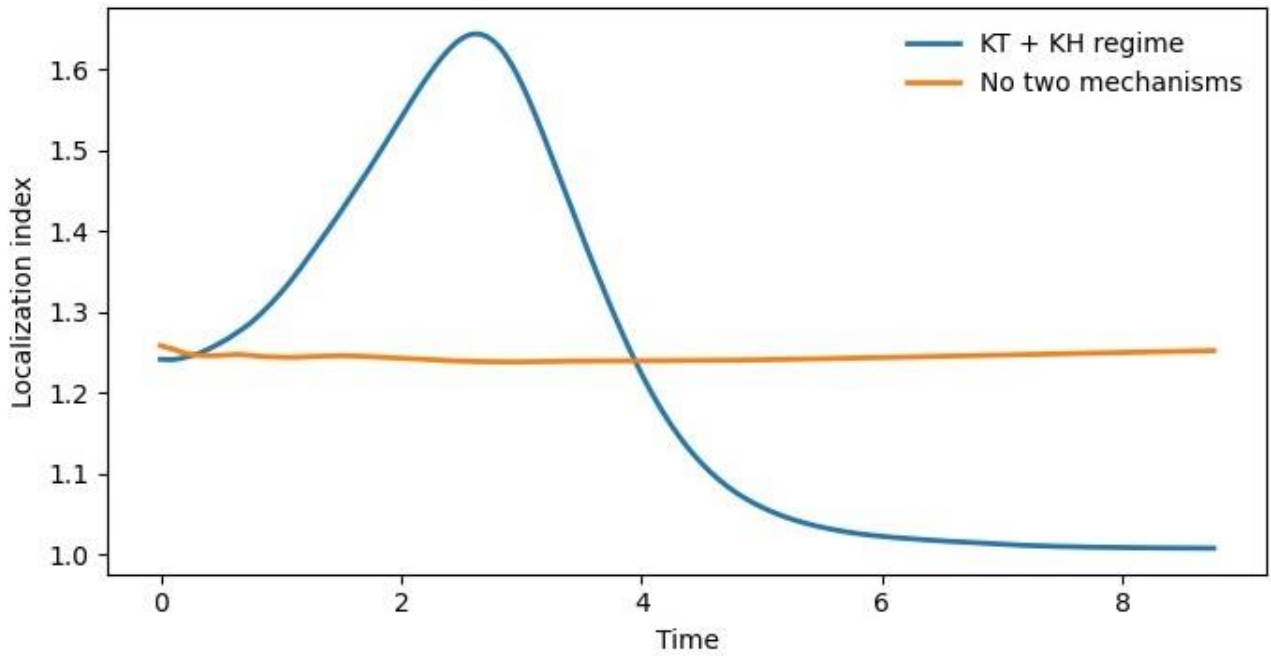

**Figure 5**. Comparison between two regimes of throat biofilm dynamics using a localization index. The index quantifies spatial concentration of interface deformation over time. In the KT + KH regime, the index increases sharply as deformation localizes around defect seeds and shear-driven amplification organizes the interface into persistent structures. In the regime lacking both defect nucleation and interfacial instability, the index remains nearly constant, indicating diffuse, unstructured dynamics.

| Feature | KT-like nucleation + KH-like shear amplification | No-two-mechanisms regime: generic growth + generic shear (no defect threshold, no instability) |
|---|---|---|
| Onset of surface structures | Abrupt appearance of localized "seeds" once a growth/hydration threshold is crossed | Gradual roughening as biomass accumulates; no sharp onset |
| Spatial localization | Hotspots at crypt rim annulus, constrictions, branch points; long-lived niches | Structures track local biomass only; no persistent hotspots beyond nutrient gradients |
| Response to airflow/saliva bursts | Nonlinear amplification: ripples/fingers appear above a shear threshold; episodic flares | Mostly proportional deformation: stronger flow gives smoother thinning or mild, reversible corrugation |
| Directionality | Patterns become flow-aligned (ripples, fingers) during high-shear episodes | Weak alignment; changes look like passive smearing |
| Space–time signature | Growth cones: deformation spreads laterally from seed sites; clear bands in space–time maps | No cones; fluctuations are diffuse, correlated mainly with overall thickness changes |
| Fragment shedding (dispersal) | Intermittent, event-triggered shedding from ripple crests/finger tips; "bursty" aerosolization risk | Continuous low-level erosion; shedding rate scales smoothly with flow |

**Table.** Comparison of the two regimes of throat biofilm dynamics.

CONCLUSIONS

We examined tonsillar biofilm dynamics by coupling localized surface heterogeneity with shear-driven interfacial organization within a simplified geometric representation of the crypt. When both mechanisms are active, interface deformation is neither spatially uniform nor transient. Instead, deformation initiates at discrete locations associated with defect sites, remains anchored to them and spreads laterally in a structured manner, producing persistent space–time bands. Growth-cone analysis demonstrated that deformation propagates outward from defect locations with well-defined fronts, while the localization index provided a quantitative measure of spatial concentration over time. In contrast, simulations performed with both mechanisms suppressed yielded diffuse fluctuations lacking spatial anchoring, propagation or persistence. Across the observables considered, including interface profiles, space–time maps and scalar metrics, the coexistence of defect nucleation and shear amplification consistently generated a distinct dynamical behavior. Therefore, within our approach, crypt geometry and mechanical heterogeneity are sufficient to impose organized structure on biofilm dynamics.

Our approach reframes tonsillar biofilms as mechanically organized systems shaped by geometry and instability rather than uniformly responding biological layers. By explicitly incorporating a Kosterlitz–Thouless–type defect nucleation mechanism, we separate the formation of persistent surface irregularities from subsequent deformation dynamics. By coupling this to a Kelvin–Helmholtz–type shear-driven instability, we distinguish nucleation from amplification, allowing each process to be associated with different physical conditions. This separation contrasts with existing models focusing exclusively on hydrodynamic instabilities or treating biofilms as spatially homogeneous viscoelastic materials (Li, Matouš, and Nerenberg 2020; Razgaleh, Wrench, and Jones 2023; Wells et al. 2023; Sun et al. 2025). Our approach provides a clearer mapping between surface heterogeneity, shear and spatial memory, without reliance on organism-specific parameters. Compared with reaction–diffusion, erosion-based or purely continuum growth models (Duddu, Chopp, and Moran 2009; Mattei et al. 2018; Stewart et al. 2019; Jin, Marshall, and Wargo 2020; Fernandes, Gomes, and Simões 2022; Klempt et al. 2024), our model prioritizes mechanistic interpretability and geometric specificity, positioning it between detailed multiphase simulations and abstract pattern-based descriptions.

Within the broader landscape of biofilm modeling strategies, our approach can be classified among the reduced, mechanism-driven models that emphasize geometry and instability as primary organizing principles. It differs from agent-based and biochemical network models by abstracting biological complexity into defect fields and effective shear terms. It also differs from purely hydrodynamic treatments, by treating persistent surface heterogeneity as a fundamental driver rather than a secondary perturbation.

Several limitations must be acknowledged. Our simulations are representative rather than predictive, relying on a toy model whose parameters were selected for numerical stability and qualitative clarity rather than empirical calibration. The Kosterlitz–Thouless–type mechanism is implemented phenomenologically, with defect unbinding imposed rather than dynamically generated. Similarly, the Kelvin–Helmholtz–type instability is represented through a reduced amplification term rather than through explicit fluid–structure interaction equations. Analyses are descriptive and based solely on simulated data, without hypothesis testing against experimental measurements. The figures illustrate regime separation and organizing principles, but should not be interpreted as quantitative predictions for biological systems.

Our approach suggests directions for future investigation. The explicit separation between defect nucleation and shear amplification implies testable hypotheses concerning the conditions under which biofilm reorganization should occur within tonsillar crypts, such as differential sensitivity to growth history versus airflow intensity. Experimental systems combining controlled surface heterogeneity with imposed shear could be used to evaluate whether localization and growth-cone–like propagation emerge as anticipated. Imaging studies might assess whether persistent deformation sites correlate with structural irregularities of crypt epithelium. From a modeling perspective, our approach could be extended by introducing dynamic defect formation, coupling the interface equation to simplified flow fields or generalizing the geometry beyond one dimension.

In summary, we addressed whether the organization of tonsillar biofilms can be understood through the interaction of localized defect formation and shear-driven instability within crypt geometry. Our results indicate that combining these two mechanisms yields a peculiar dynamical regime characterized by localization, propagation and spatial memory. Our approach suggests that mechanical structure and geometry could impose order on biofilm dynamics independently of biological detail, providing a physical lens through which persistence and episodic reorganization could be jointly interpreted.

## DECLARATIONS

**Ethics approval and consent to participate.** This research does not contain any studies with human participants or animals performed by the Author.

**Consent for publication.** The Author transfers all copyright ownership, in the event the work is published. The undersigned author warrants that the article is original, does not infringe on any copyright or other proprietary right of any third part, is not under consideration by another journal and has not been previously published.

**Availability of data and materials.** All data and materials generated or analyzed during this study are included in the manuscript. The Author had full access to all the data in the study and took responsibility for the integrity of the data and the accuracy of the data analysis.

**Competing interests.** The Author does not have any known or potential conflict of interest including any financial, personal or other relationships with other people or organizations within three years of beginning the submitted work that could inappropriately influence or be perceived to influence their work.

**Funding.** This research did not receive any specific grant from funding agencies in the public, commercial or not-for-profit sectors.

**Acknowledgements:** none.

**Authors' contributions.** The Author performed: study concept and design, acquisition of data, analysis and interpretation of data, drafting of the manuscript, critical revision of the manuscript for important intellectual content, statistical analysis, obtained funding, administrative, technical and material support, study supervision.

**Declaration of generative AI and AI-assisted technologies in the writing process.** During the preparation of this work, the author used ChatGPT 4o to assist with data analysis and manuscript drafting and to improve spelling, grammar and general editing. After using this tool, the author reviewed and edited the content as needed, taking full responsibility for the content of the publication.